\documentclass[aps,prl,twocolumn,superscriptaddress]{revtex4-1}
\usepackage{amsmath}
\usepackage{graphicx}
\usepackage{braket}
\usepackage{soul}

\usepackage[usenames, dvipsnames]{color}
\graphicspath{{figures/}{}}
\begin{document}

\title{Development of transmon qubits solely \\
from optical lithography on 300mm wafers}


\author{N. Foroozani}
\affiliation{Laboratory for Physical Sciences, University of Maryland, College Park, MD 20740, USA}
\affiliation{Department of Physics, University of Maryland, College Park, MD 20742, USA}
\author{C. Hobbs}
\affiliation{SUNY Polytechnic Institute, Albany, NY 12203, USA}
\author{C. C. Hung}
\affiliation{Laboratory for Physical Sciences, University of Maryland, College Park, MD 20740, USA}
\affiliation{Department of Physics, University of Maryland, College Park, MD 20742, USA}

\author{S. Olson}
\affiliation{SUNY Polytechnic Institute, Albany, NY 12203, USA}
\author{D. Ashworth}
\affiliation{SUNY Polytechnic Institute, Albany, NY 12203, USA}
\author{ E. Holland}
\affiliation{SUNY Polytechnic Institute, Albany, NY 12203, USA}
\author{ M. Malloy}
\affiliation{SUNY Polytechnic Institute, Albany, NY 12203, USA}
\author{P. Kearney}
\affiliation{SUNY Polytechnic Institute, Albany, NY 12203, USA}
\author{B. O'Brien}
\affiliation{SUNY Polytechnic Institute, Albany, NY 12203, USA}
\author{ B. Bunday}
\affiliation{SUNY Polytechnic Institute, Albany, NY 12203, USA}
\author{D. DiPaola}
\affiliation{SUNY Polytechnic Institute, Albany, NY 12203, USA}
\author{W. Advocate}
\affiliation{SUNY Polytechnic Institute, Albany, NY 12203, USA}
\author{T. Murray}
\affiliation{SUNY Polytechnic Institute, Albany, NY 12203, USA}
\author{P. Hansen}
\affiliation{SUNY Polytechnic Institute, Albany, NY 12203, USA}
\author{S. Novak}
\affiliation{SUNY Polytechnic Institute, Albany, NY 12203, USA}
\author{S. Bennett}
\affiliation{SUNY Polytechnic Institute, Albany, NY 12203, USA}
\author{M. Rodgers}
\affiliation{SUNY Polytechnic Institute, Albany, NY 12203, USA}
\author{B. Baker-O'Neal}
\affiliation{SUNY Polytechnic Institute, Albany, NY 12203, USA}
\author{B. Sapp}
\affiliation{SUNY Polytechnic Institute, Albany, NY 12203, USA}
\author{E. Barth}
\affiliation{SEMATECH, Albany, NY 12203, USA}
\author{J. Hedrick}
\affiliation{SUNY Polytechnic Institute, Albany, NY 12203, USA}
\author{R. Goldblatt}
\affiliation{SUNY Polytechnic Institute, Albany, NY 12203, USA}
\author{S. S. Papa Rao}
\affiliation{SUNY Polytechnic Institute, Albany, NY 12203, USA}
\author{K. D. Osborn}
\affiliation{Laboratory for Physical Sciences, University of Maryland, College Park, MD 20740, USA}
\affiliation{Joint Quantum Institute, University of Maryland, College Park, MD 20742, USA}


\date{\today}

\begin{abstract}

Qubit information processors are increasing in footprint but currently rely on e-beam lithography for patterning the required Josephson junctions (JJs). Advanced optical lithography is an alternative patterning method, and we report on the development of transmon qubits patterned solely with optical lithography. The lithography uses 193 nm wavelength exposure and 300-mm large silicon wafers. Qubits and arrays of evaluation JJs were patterned with process control which resulted in narrow feature distributions: a standard deviation of $0.78\%$ for a 220 nm linewidth pattern realized across over half the width of the wafers. Room temperature evaluation found a $2.8-3.6\%$ standard deviation in JJ resistance in completed chips. The qubits used aluminum and titanium nitride films on silicon substrates without substantial silicon etching. $T_1$ times of the qubits were extracted at 26 $\mu s$ -- 27 $\mu s$, indicating a low level of material-based qubit defects. This study shows that large wafer optical lithography on silicon is adequate for high-quality transmon qubits, and shows a promising path for improving many-qubit processors.

\end{abstract}

\pacs{}

\maketitle

\section{I. Introduction}
Currently, quantum processors allow demonstrations of quantum error correction \cite{Pfaff, IBM2, IBM3, martinis}, chemical simulations \cite{neda, Malley}, and factoring \cite{martinis3}. These prototype processors contain at least four transmon qubits \cite{Wang, fourQubits} as high coherence elements on small (e.g., 5 mm) chips \cite{IBM3, martinis3}. There have been reports of these superconducting qubits, fabricated on sapphire substrates and measured in 3D cavities, with relaxation ($T_1$) and coherence times of 100 $\mu s$ \cite{Wang, TransmononSapphire}. On silicon substrates, the standard for CMOS fabrication, the same transmon qubits generally have shorter $T_1$ times, with an optimized design showing approximately 30 (50) $\mu s$ with (without) a silicon substrate etch to remove material-based loss \cite{bridgeQubit}. In an attempt to increase the size of quantum processors, larger (e.g., 10 mm) chips are being optimized to reduce the affects from parasitic microwave modes in the sample box and wiring crosstalk \cite{wirebond, hiddenmodes, Bronn}.

%
While coherent qubits and bus resonators have a large footprint, the former require a small Josephson junction with a tunneling barrier area of $\leq0.04$  $\mu$m$^2$ and a lateral JJ dimension of $\leq 200$ nm \cite{PerDelsingWnote, Frunzio, Hanhee, Wang, Gambetta2017}. While all JJs are believed to have two-level defects in the tunneling barrier that resonantly coupled to the qubit, small junctions may result in enhanced coherence because the contained defect density per unit frequency should be smaller than in large JJs \cite{Martinis2005, Stoutimore2012}. Statistics of these small JJs have been collected using $\geq$ 100 JJs on small chip areas \cite{stable, PerDelsingWnote}. In the first study \cite{stable} the write field is only apparently 2 mm in its largest dimension and intentionally set to be small for uniformity. There it was found that the normal-state resistance of these JJs has a standard deviation of 3.5$\%$ for all JJs in one fabrication run, a statistic which is key since it is inversely proportional to the Josephson critical current. In the second study \cite{PerDelsingWnote} many dc-SQUIDs were evaluated on few-mm chips. They were found to have a lower standard deviation (for pairs of JJs within) in this small field area. The required tolerance for JJ critical current and related qubit frequency depends on the quantum computing gate architecture. In a successful non-tunable architecture, cross-resonant gates are used on the qubits \cite{chow, McKay, Versluis}. In this technique the qubits must have a frequency interval close to the qubit anharmonicity, which is only 3-5$\%$ of the qubit frequency, and places challenging constraints on the lithography. While e-beam lithography works well for previous circuits, it also appears that this method is unproven for large exposure fields (e.g., 10 mm quantum processors). 

Optical lithography, developed in the CMOS industry, is well known to produce nearly identical features over large areas \cite{Harriott}. The narrowest resolvable lithographic line is given by $k_{1} \lambda/$$\it{NA}$, indicating the important variables of lithographic wavelength $\lambda$ and numerical aperture $\it{NA}$. A numerical factor close to unity, $k_1$, accounts for  process-related parameters such as resolution-enhancement techniques, coherence of the light source, etc \cite{Pedrotti}. Following the introduction of 365 nm wavelength lithography (known as i-line), came 248 nm wavelength (known as Deep UV) started producing CMOS processors in 1999 \cite{Takashi, Harriott}. While we are not aware of published reports of transmons fabricated solely from optical lithography, there is a recent set of studies on 248 nm optical lithography for JJ digital logic circuits \cite{Sergey, Sergey2, Sergey3}. The first study found a standard deviation of $8\%$ for the resistance of 300 nm diameter Nb/AlOx JJs on chips from 248 nm lithography on 200 mm wafers \cite{Sergey2}. Furthermore, a need for 248 nm and 193 nm lithography was identified for JJ circuits \cite{Sergey2}. 

Here we present the first report of transmon qubits patterned solely with optical lithography on 300 mm wafers at the facility at SUNY Polytechnic. This facility is already used for integrated photonics and neuromorphic computing research \cite{Pops1, Pops2, Pops3, Pops4, Pops5}. These qubits depend on 220 nm linewidth features for JJs fabricated using 193 nm photo lithography on 300 mm silicon wafers. The resist patterning was characterized over a $\sim$ 170 mm region of the wafer. JJs, fabricated in test arrays and qubits, were completed on coupons (chips) with a 33 mm $\times$ 25 mm area, with a small standard deviation in resistance. 
Two nominally-identical qubits separated by 12.5 mm were characterized in their relaxation, and coherence times. The study reveals the realization of high-performance qubits fabricated using advanced tools developed for the CMOS-industry, i.e., large-field optical lithography on silicon substrates. Optical lithography, using a reticle patterned as shown in Figure 1, permits Josephson structures to be `written' into the photoresist in parallel on many large chips. In contrast, e-beam lithography actually writes the pattern serially with a beam, and thus is relatively slow for large patterns. Furthermore it must use stitching of many write fields to realize 10 mm chip.

\section{II. Fabrication}
While a wide range of resistivity, up to 20 k$\Omega$-cm, is available on smaller Si wafers, it is more limited for 300 mm Si wafers leading to the question of whether high performance qubits can be realized on the latter wafers. This study utilizes two batches with resistivity specified in the range of 2-13 k$\Omega$-cm. We use the batch with larger resistivity of 10 k$\Omega$-cm for millikelvin measurements of qubits. Measurements using a KLA-Tencor SP3 detect only 10 and 17 particulate defects $>$90 nm from two randomly selected wafers, as evidence of starting surface cleanliness of the wafer. 40nm of TiN is grown using physical vapor deposition after a dilute HF-last surface clean of the Si wafer, and found to have a (200) texture by XRD analysis. The TiN film is then patterned to form the probe pads in the Josephson junction arrays and the paddles in the qubits (see Fig. 2(a)). It should be noted that the silicon surface outside the TiN pattern will be once again covered with its native oxide. A wafer is then spin-coated with 530 nm of polydimethylglutarimide-based resist (PMGI SF6 series, from MicroChem) in a stand-alone track, and baked. This is followed by spin-coating (and baking) a 208 nm-thick layer of a commercial 193 nm positive photoresist. The pattern for the junctions is transferred into this resist layer using an ASML TWINSCAN AT:1200B system. 

\begin{figure}
\includegraphics[width=0.46\textwidth]{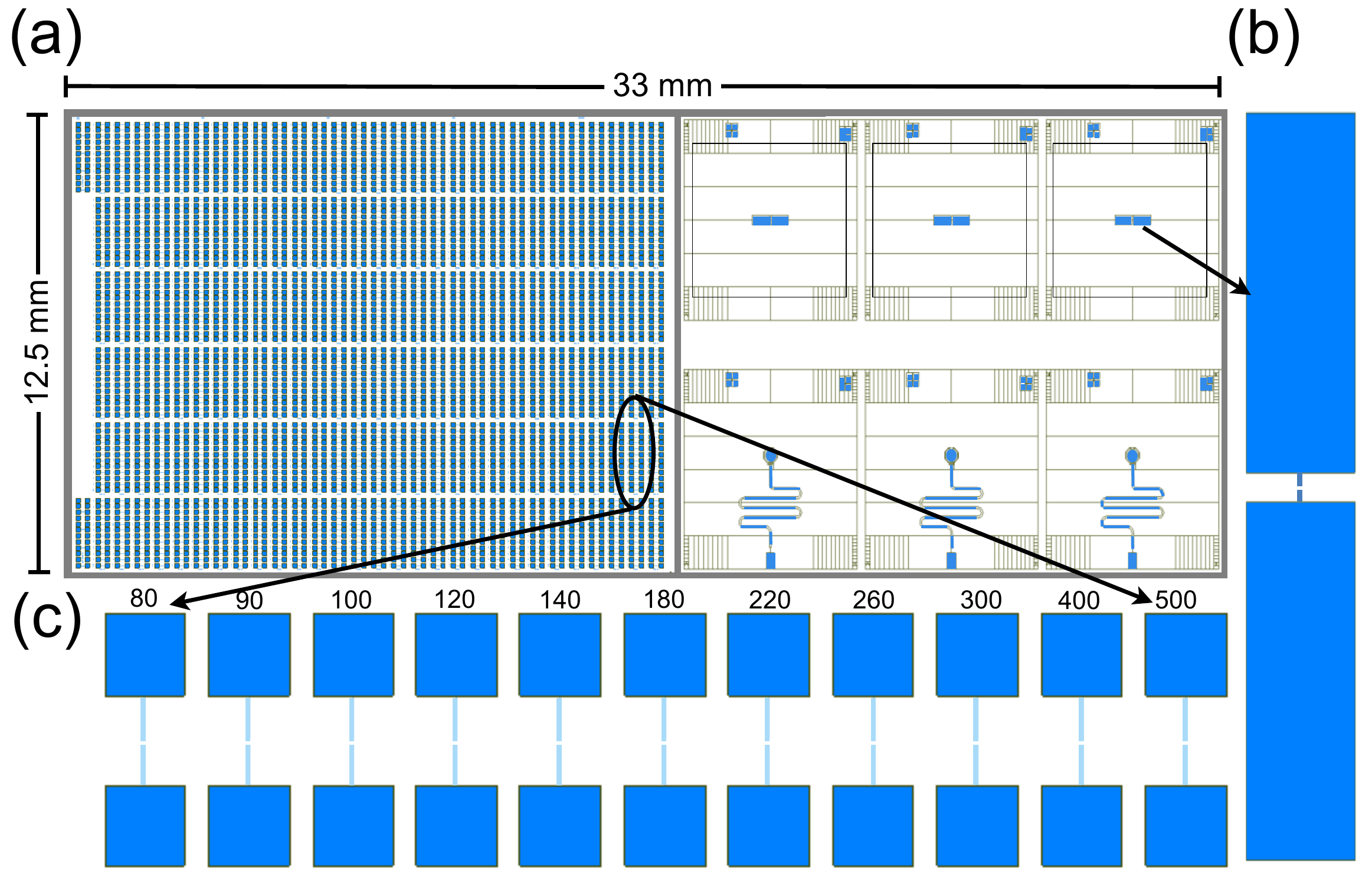}
\caption{(a) The mask pattern used for 193 nm-resist lithography across 300 mm wafers. In this work resistance is generally probed on a coupon (chip) contains 2 copies of the mask pattern for a JJ sampling area of approximately 25 mm $\times$ 16 mm, and 48 JJs for each finger width and bridge width. The upper right of (a) contains 3D transmons which are eventually diced into 5 mm chips (square shown), where one 3D transmon pattern shown in (b). The left part of the pattern (a) contains JJ arrays for room temperature probing. For example, (c) shows pattern for one Dolan-bridge width and different (dimension in nm) JJ finger widths. The lower right patterns were not measured for this work.}
\end{figure}

The pattern is 33 mm $\times$ 12.5 mm, which is repeated across the wafer and containing two sections: arrays of individual JJs and qubits (see Fig. 1). After exposure the wafer is immediately baked and developed using a 0.26N TMAH-based solution. Through optimization of development time, resist is not only selectively removed to form the Dolan bridge, but the amount of underlying lift-off resist is also removed from beneath the bridge (see Fig. (2b)). The latter is confirmed using tilt-view SEM observations of several samples. Josephson junction arrays with varying junction dimensions are formed with top electrode `finger' widths ranging from 80 nm to 500 nm (see Fig. 1(c) and Fig. 2(a) (iv)), and Dolan bridge widths ranging from 380 nm to 500 nm. The 220 nm finger feature in the array is  targeted for lithographic exposure optimization since the qubits in this study used 220 nm wide finger feature for JJ formation (and incidentally use a Dolan bridge width of 440 nm). Figure 2(f) shows the measured finger feature widths across a central 176 mm $\times$ 130 mm region of the wafer, associated with a standard deviation of 1.7 nm, i.e., 0.78$\%$ of the median. Measurements of our 440 nm wide bridges in the array are confirmed to have similarly low standard deviation, 0.67$\%$ of the median.

The wafer is then cleaved into coupons (chips) of approximately 45 mm $\times$ 35 mm area containing at least 2 repeated patterns (see Fig. (1)). The coupons are loaded into an e-beam evaporation system that is `pre-seasoned' by two evaporations of aluminum. Lack of metal contamination from prior users of the evaporation tool is verified through secondary-ion mass spectroscopy (SIMS) analysis of the deposited aluminum films.
Starting from a pressure below 5$\times$10$^{-7} $ Torr, the wafer is then sputtered with Ar for 4 minutes to clean organic contaminants. 
Separate tests determine that the sputtering decreases photoresist film thickness but not SiO$_2$ film thickness. After sputter-cleaning for $\approx$ 60 s in base pressure, the bottom electrode of the JJ is formed by normal ($0^{\circ}$ sample tilt) evaporation of 30 nm of aluminum, at a deposition rate of 0.1 nm/s. The aluminum is subjected to in-situ oxidation with a $20\%$ O$_2$ in Ar mixture flowing at a controlled rate of 800 sccm, with the turbomolecular pump closed off and the chamber connected to the roughing pump. The system pressure under these flow conditions is $\approx$ 0.23 Torr based on measurements during several nominally identical runs and oxidation times fall in the range 30--120 minutes (to tune the junction resistance). After completing oxidation, we pump the chamber down and deposit 60 nm of Al using $55^{\circ}$ evaporation at the previous deposition rate (see Fig. 2(c)). The Al deposition made contact to the underlying TiN paddles. The coupons are subjected to lift-off with NMP with gentle mechanical agitation, followed by an IPA rinse and N$_2$ blow-dry (see Fig. 2(d)). Figure 2(d) and Ref. \cite{ECS} indicate that the JJ area had variability beyond the tightly controlled linewidth in the resist pattern caused by rough evaporated-line features, and we view this as an important contributor to the variability of the realized JJ area. Transmission electron microscopy of cross-sections of the JJ confirms good step coverage and continuity of the top Al layer over the bottom Al layer as well as the presence of a thin layer of native oxide of Si under the Al as expected (see Fig. 2(e)). This oxide could limit the relaxation time of qubits due to dielectric loss \cite {Sarabi, Lisenfeld, Wang} (but can be addressed in the future).

\begin{figure}
\includegraphics[width=0.44\textwidth]{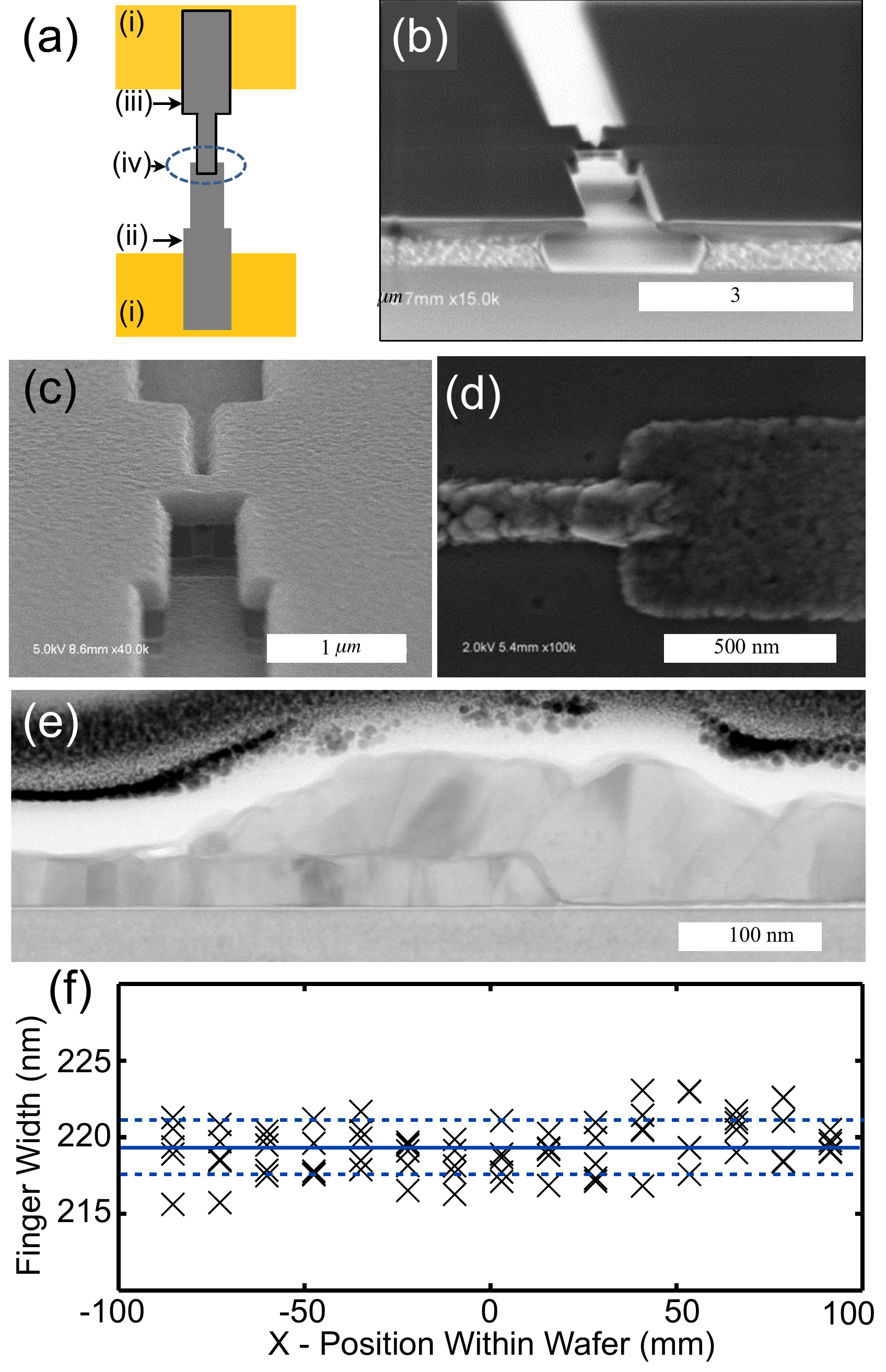}
\caption{(a) Patterns formed at various stages of the process flow: (i) TiN paddles, (ii $\&$ iii) leads and (iv) (in dashed circle) connected Josephson junction which is formed by two angled evaporations and oxidation near the Dolan bridge. Top Al evaporation finger overlaps an oxidized Al base layer. (b-d) SEM micrographs at consecutive stages (b) Patterned Dolan bridge before evaporations (c) bridge with JJ (d) JJ after liftoff (bridge removed).(e) cross-sectional TEM image of angle-evaporated JJ. (f) Inline SEM measurements of finger width in resist, with fit to mean (solid line) and $\pm$ standard deviations $\sigma$ (dashed line). Finger widths of top Josephson electrode with standard deviation of $0.78\%$. The Dolan bridge width has a similar standard deviation to mean ratio (see text).}
\end{figure}

\section{III. measurement}
\subsection{A. Room-Temperature Characterization}
We characterize JJ arrays using a voltage sweep from -0.2V to 0.2V. We extract JJ resistance using a linear fit to I-V curves. Data revealing open, shorted, or nonlinear I-V curves (with a coefficient of determination $R^2$ $<$ 0.999), are defined as `bad' and excluded from further analysis. The yield of `good' junctions in the arrays is well over $90\%$ for finger widths of 140 nm and greater. The JJ resistance variation from the mean for 3 fabrication runs in shown in Fig. 3(a) for the oxidation process that yields $\approx$ 5 k$\Omega$ in target JJs: those of 220 nm finger width and 440 nm bridge width. Each variation data point (for a given finger width) used a set of 48 sampled JJs spread over the area of $\approx$ 25 mm$\times$16 mm on the chip. For the targeted finger width, which is 220 nm, we found that the resistances had standard deviations of 2.8$\%-3.6\%$. Our variability compares well to optimized e-beam lithography which was sampled over a much smaller area \cite{stable}. On each chip with the same JJ arrays as above, there are two qubits with the target JJ type, spaced by 12.5 mm from one another (recall the half chip pattern in Fig. 1(a)). These qubits were probed on four chips with generally different oxidation conditions. From these data, the hand-probed qubit JJ resistance was found to be within the $95\%$ confidence bounds of the corresponding JJs in the arrays, as expected (see Fig. 3(b)). Chip 1 and 2 were nominally oxidized the same way, and the 8.5$\%$ difference between these runs is very close to the run-to-run difference seen previously for similarly small JJs \cite{PerDelsingWnote}. Chips 3 and 4 were intentionally fabricated with a larger oxidation time. 

Now we compare our standard deviation of 2.8$\%-3.6\%$ for an area of $\approx$ 25 mm$\times$16 mm with other fabrication techniques. E-beam lithography is known to allow uniform JJs with a standard deviation of 3.5$\%$ over an approximate 2 mm distance. This is a similar variation but the JJs are located in an order-of-magnitude smaller lateral distance than this study. The previous deep sub-micron optical JJ fabrication for digital circuits \cite{Sergey2} used $\lambda$ = 248 nm wavelength on 200 mm wafers, and contrasts our $\lambda$=193 nm process on 300 mm wafers in a number of ways. Their fabrication exposed a circular area for the JJ, whereas ours exposed a line ending in a gap for the Dolan bridge. In their 800 nm diameter JJs the standard deviation is 3$\%$ in resistance, and for the 300 nm diameter JJs the standard deviation is 8$\%$. Our standard deviation for the 220 nm line JJ is 2.8$\%-3.6\%$. Although our 220 nm feature JJs show lower variability compared to their 300 nm diameter JJs, the fabrication techniques are also different. 200 nm lines 
have recently been fabricated for superconducting circuit elements using 248 nm lithography \cite{Sergey3}. Some of the excess JJ resistance standard deviation of $\sim$3$\%$ over the photolithograph linewidth deviation 0.78$\%$, is due to uniformity limited by the Dolan-bridge fabrication; other techniques are known which define the JJ only with straight edges and hence might improve upon this \cite{pappas_overlap_jj}.    
\begin{figure}
\includegraphics[width=0.39\textwidth]{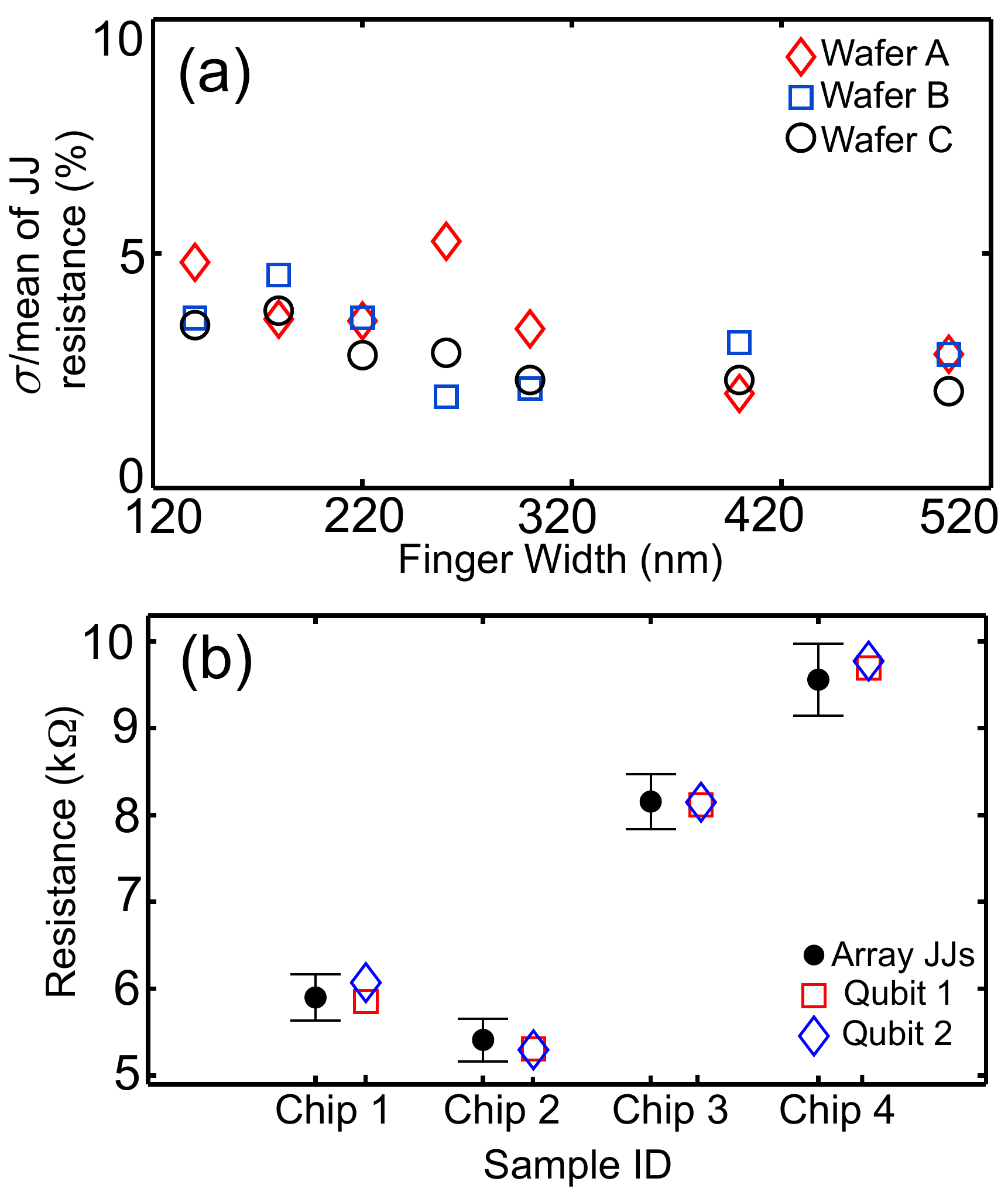}
\caption{(a) Standard deviation of Josephson junction resistance divided by median value, as a function of finger width for 3 wafers. (b) Mean resistance (black circle) with $95\%$ confidence interval (black bars) of 220 nm-nominal finger width JJs (48), along with corresponding qubits on a 33 mm $\times$ 25 mm area of a chip (see Fig. 1). Different chips were generally oxidized with intentionally different conditions (see text).The qubit resistance values are within the measured ranges of the corresponding JJ arrays, indicating uniform results across the chip.}
\label{power_saturation}
\end{figure}

\subsection{B. Millikelvin Characterization}
Two transmon qubits from 12.6 mm separation on the wafer are measured. Each transmon is nominally the same, with a paddle spacing of 40 $\mu$m, paddle width of 250 $\mu$m, paddle length of 500 $\mu$m, for a total transmon length of 1.040 mm (see Fig. 1(b)). Each qubit, on a 5 mm square chip, is mounted in its own 3D aluminum cavity with two measurement ports, creating qubit-cavity samples 1 and 2 (see Fig. 4 (a)). The samples are both cooled to 10 mK in a cryogen-free dilution refrigerator. The output signal from the cavity, passing through a 12 GHz low pass filter and three microwave isolators, is amplified at 3K using a low-noise HEMT amplifier. After further amplification at room temperature, two quadrature signals of the resonator are acquired through homodyne detection and then digitized after low-pass filtering. 
For samples 1 (shown in Fig 4 (b)) and 2, the large-drive-power or bare resonance frequency occurs at $\omega_c/2\pi=7.8291$ GHz and 7.7517 GHz, respectively. Using the qubit induced dispersive shift of the cavity at low power, we extracted the dispersive shift of $\chi/2\pi=-0.86$ MHz and --0.97 MHz, respectively \cite{Blais}. The cavity linewidths revealed cavity decay rates of $\kappa/2\pi=$ 340 kHz and 370 kHz (for sample 1 and 2).

\begin{figure}
\includegraphics[width=0.48\textwidth]{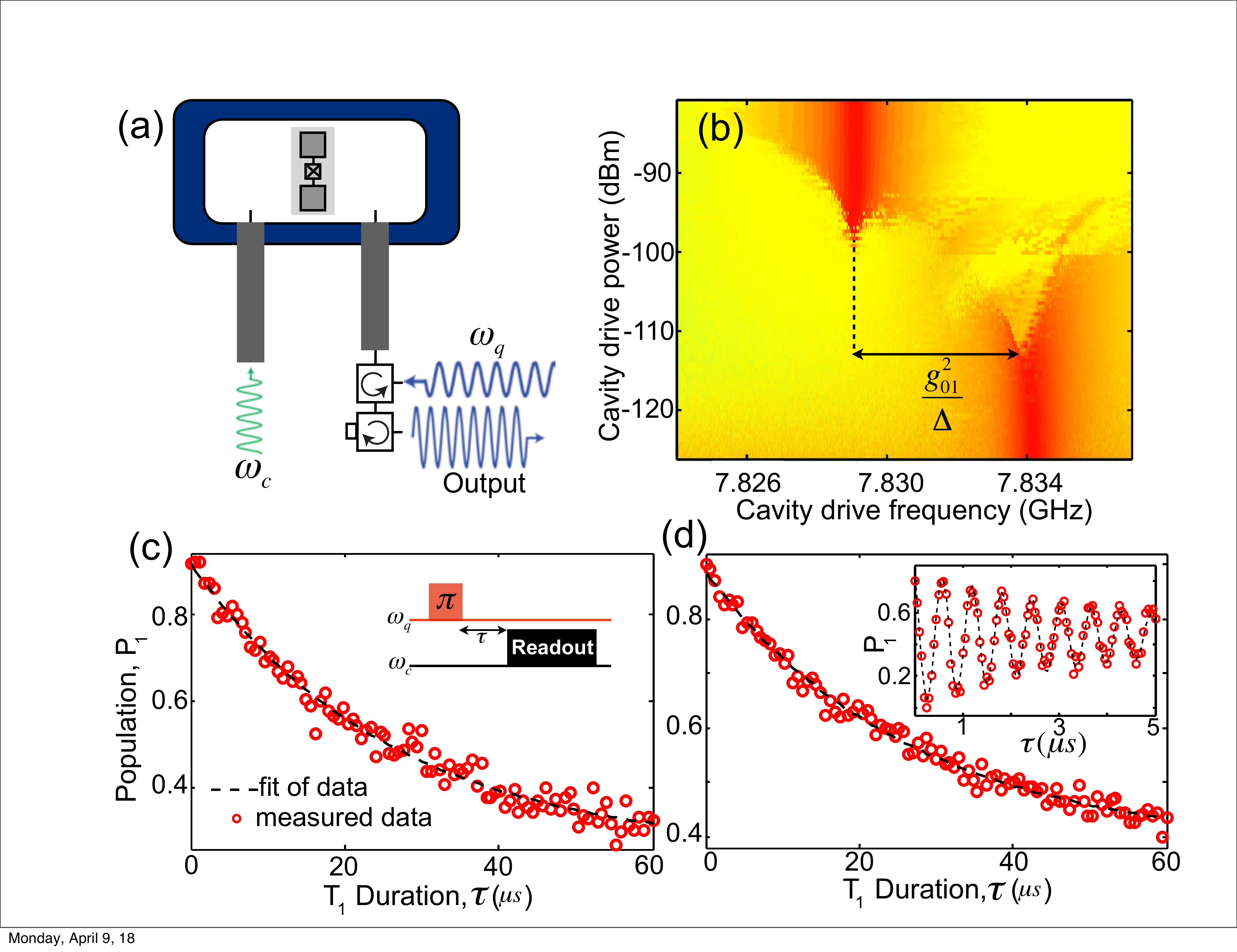} 
\caption{(a) Qubit and cavity diagram showing microwave measurement ports and frequencies $\omega_q$ and $\omega_c$. (b) Cavity transmission as a function of drive frequency and drive power. The response above -98 dBm is due to the bare cavity. For low drive strength, the cavity is shifted by $\chi_{01}=g_{01}^2/\Delta$. For readout we use a high power technique to quickly project the qubit and cavity into a classically distinguishable system using power at $f_{bare}$. The optimal power is found when the maximum difference in transmission is found for the qubit initially in the ground and excited states \cite{Reed}. (c) and (d) Relaxation data on qubit 1 and 2, respectively. (d inset) Ramsey data on qubit 2. }   
 \label{optical image}
\end{figure}

We perform qubit spectroscopy and more precise measurements using Ramsey fringes to find the qubit frequencies. The results show the $1.33\%$ difference in qubit transition frequencies. Additionally, using excitations to the second excited state we obtain charging energies $E_C/h$ of approximately 290 and 280 MHz for samples 1 and 2, respectively. 
From measurements of $\chi=g^2/\Delta$ and qubit-cavity detuning $\Delta$, we extract couplings of approximately $g/2\pi$ = 54 MHz for both qubits. 
The transition frequency of the transmon can be expressed as $\omega_q= \sqrt{\pi \Delta/(\hbar R_N  C)}$, where $R_{N}$ is the junction resistance and C is the total capacitance. Since over $>$ 90$\%$ of C is attributed to fields external to the Josephson junction trilayer, i.e., the paddles, shunt capacitance, C is relatively constant. This explains why the $E_C$ is the same between qubits within the precision of our measurements. Using this square root scaling of the qubit frequency and the above standard deviation for the room-temperature resistance ($2.8-3.6\%$), we find that this corresponds to an expected standard deviation of $\leq1.8\%$ in qubit frequencies. As expected the $1.33\%$ difference observed in qubit frequencies is consistent with the room temperature JJ characterization.

In time domain measurements, Rabi oscillations of the qubit are obtained to calibrate the pulse amplitude with a $\pi$-pulse duration of approximately 60 ns. We characterize the qubit's energy relaxation by performing a $T_{1}$ measurement, which is often a limiting factor in coherence time. First we apply a $\pi$ pulse precisely at the qubit transition frequency, and then wait for a variable amount of time, $\tau$, to apply the readout pulse \cite {Reed}. The decay of population in the excited state, $P_1$ is shown in Fig. 4 (c) and (d). Each data point represents the average measured voltage of $10^4$ identical experiments for a given value of $\tau$. The data is fitted by a single exponential for an extracted $T_1$ of 26.1 and 26.7 $\mu$s in qubits 1 and 2, respectively. These times were measured over multiple days and deemed to be limited by loss since the Purcell decay is negligible. The $T_1$ value is comparable with state-of-the-art transmon design on Si which has a relaxation time of approximately 30 $\mu s$ \cite{bridgeQubit} prior to silicon etching (and 
JJ suspension) which reduces the presence of materials. We note that other work uses deep etching with nitride superconductors for high coherence \cite{Bruno}, while our etching is only on the nanometer scale. Our work thus finds a low level of loss from carefully prepared materials: the large-Si substrate, the PVD TiN, evaporated Al, and material interfaces with some native oxides.

We also performed a Ramsey measurement to determine qubit frequency precisely and dephasing time $T^{*}_2$ (see inset of Fig. 4(d)). For this, the measurement sequence consisted of two $\pi/2$ pulses to the qubit at a frequency 1-2 MHz detuned from the expected qubit frequency with a varying time delay $\tau$ between the pulses, followed by a readout pulse. The Ramsey measurement gives fringes of the excited state population oscillating in time. The oscillation frequency is precisely the difference between the qubit frequency and the drive frequency, which we use to obtain the qubit transition frequencies 4.7661 GHz and 4.7027 GHz, respectively for sample 1 and 2. Decreasing contrast in the fringes corresponds to the loss of phase coherence with time. The fit to an exponential sinusoidal decay of the fringes revealed the $T^{*}_2$ of 4.4 and 4.9 $\mu s$ for qubit 1 and 2, respectively and we show the latter in the inset to Fig. 4(d). It is expected that this low value of $T^{*}_2$ is limited by thermal noise injected into our measurement setup \cite{Palmer}. 

\section{VI. Conclusion and Outlook}
In conclusion, in this study we find a method to fabricate high quality transmons using advanced optical lithography. Our JJs are formed over a field which is approximately an order of magnitude larger than JJs made with the previous method of fabrication, which is e-beam lithography, but they have similar variability. Additionally, our technique uses 193 nm lithography from 300 mm wafers, and  should ultimately provide benefits over 248  nm lithography on 200 mm wafers, although the JJs fabricated with the latter are fabricated differently. A standard deviation of only 0.78$\%$ from 220 nm width line is found across the central 176 mm $\times$ 130 mm on 300 mm wafers. JJ statistical tests where performed on 25 mm $\times$ 16 mm fields. For the JJ array portion of these chips the standard deviation in resistance is $2.8-3.6\%$.
Two qubits were studied in the time domain. They had a difference in 0-1 transition frequency of $1.33\%$, a value consistent with the JJ statistics on resistance. Furthermore, they exhibit relaxation times of $T_1$= 26--27 $\mu s$. We conclude that the defects in the bulk Si and fabricated films are comparable to state-of-the-art transmons on silicon. In the future we plan to extend the coherence time to over 100 $\mu s$ while demonstrating further uniformity improvements in JJs for qubits.
\section{acknowledgments}
The authors at SUNY Polytechnic are grateful to Jon Mcmahon, David Eason and members of his team at the Univ. at Buffalo, NY for use of the Shared Instrument Lab. The SUNY Poly team is particularly indebted to Britton Plourde (Syracuse University) and to David Pappas (NIST Boulder) for guidance on qubit design and testing, as well as for screening measurements of qubits at their respective facilities.

\end{document}